\documentclass[conference]{IEEEtran}
\IEEEoverridecommandlockouts
\usepackage{cite}
\usepackage{amsmath,amssymb,amsfonts}
\usepackage{algorithmic}
\usepackage{graphicx}
\usepackage{textcomp}
\usepackage{xcolor}
\usepackage{booktabs}
\usepackage{lipsum}
\usepackage{multirow}
\usepackage{caption}
\usepackage{subcaption}
\usepackage{orcidlink}
\usepackage{csquotes}
\usepackage{changes}
\def\BibTeX{{\rm B\kern-.05em{\sc i\kern-.025em b}\kern-.08em
    T\kern-.1667em\lower.7ex\hbox{E}\kern-.125emX}}
\usepackage{mathtools}
\DeclarePairedDelimiterX{\infdivx}[2]{(}{)}{%
  #1\;\delimsize\|\;#2%
}

\usepackage{hyperref}

\usepackage{todonotes}           

\usepackage{lipsum}
\usepackage{fancyhdr}
\fancypagestyle{sec}{\lhead{\tmpx}}

\fancypagestyle{arXiv}
{
   \fancyhf{}
   \chead{\textcolor{gray}{This paper has been accepted to be presented at the 2025 International Symposium on Circuits and Systems (ISCAS), \\London, UK, May 25–28, 2025}}
   \cfoot{ \fontsize{=9pt}{10pt}\selectfont  \textcolor{gray}{© 2025 IEEE. Personal use of this material is permitted. Permission from IEEE must be obtained for all other uses, in any current or future media, including reprinting/republishing this material for advertising or promotional purposes, creating new collective works, for resale or redistribution to servers or lists, or reuse of any copyrighted component of this work in other works}\\\fontsize{=10pt}{12pt}\selectfont\thepage}
   
}

\begin{document}

\title{CleanUMamba: A Compact Mamba Network for Speech Denoising using Channel Pruning\\
}

\author{ 
Sjoerd~Groot,
Qinyu~Chen,
Jan C. van Gemert,
Chang~Gao

\thanks{Corresponding Author: Chang Gao (chang.gao@tudelft.nl)}
\thanks{S. Groot, J. C. van Gemert and C. Gao are with the Faculty of Electrical Engineering, Mathematics and Computer Science, Delft University of Technology, The Netherlands.}
\thanks{Q. Chen is with the Leiden Institute of Advanced Computer Science (LIACS), Leiden University, The Netherlands.}
}

\maketitle

\begin{abstract}
This paper presents \texttt{CleanUMamba}, a time-domain neural network architecture designed for real-time causal audio denoising directly applied to raw waveforms. \texttt{CleanUMamba} leverages a U-Net encoder-decoder structure, incorporating the Mamba state-space model in the bottleneck layer. By replacing conventional self-attention and LSTM mechanisms with Mamba, our architecture offers superior denoising performance while maintaining a constant memory footprint, enabling streaming operation. To enhance efficiency, we applied structured channel pruning, achieving an 8X reduction in model size without compromising audio quality. Our model demonstrates strong results in the Interspeech 2020 Deep Noise Suppression challenge. Specifically, \texttt{CleanUMamba} achieves a PESQ score of 2.42 and STOI of 95.1\% with only 442K parameters and 468M MACs, matching or outperforming larger models in real-time performance. Code will be available at: \url{https://github.com/lab-emi/CleanUMamba}
\end{abstract}


\section{Introduction}
\thispagestyle{arXiv}
Audio denoising, or speech enhancement, removes background noise from speech recordings while preserving quality and intelligibility. This technology is important for applications such as hearing aids, audio calls, and speech recognition systems. Traditional methods like spectral subtraction~\cite{boll1979suppression} and Wiener filtering~\cite{lim1979enhancement} struggle in dynamic environments, especially when noise overlaps with speech.

Deep neural networks (DNNs) have significantly advanced this field in recent years. Various architectures, including convolutional neural networks (CNNs)~\cite{pascual2017segan, luo2019conv, stoller2018wave, pandey2019tcnn}, recurrent neural networks (RNNs)~\cite{Défossez_2020_demucs, braun2021towards, cheng24_interspeech}, and transformers~\cite{Kong_2022_CleanUNet, wang2023tf}, have been used to enhance speech. Although transformers achieve high-quality results, they are computationally expensive at longer input sequences.

The Mamba state space model~\cite{Gu_2023_Mamba} offers a promising solution for sequence modeling and time series prediction. Mamba enables parallel computation during training and recurrent processing during inference without sequence length constraints, making it a strong candidate for audio denoising.

In this work, we introduce \texttt{CleanUMamba}, a neural network designed for real-time audio denoising that processes raw waveforms. \texttt{CleanUMamba} adopts the U-Net encoder-decoder architecture from~\cite{Kong_2022_CleanUNet}, replacing self-attention with Mamba state-space blocks, which enables a more compact model with reduced algorithmic latency. Additionally, we implement structured pruning to reduce the model size while maintaining high performance.

Our main contributions are:
\begin{enumerate}
    \item A novel Mamba-based architecture for time-domain speech enhancement with a 12 ms real-time algorithmic latency.
    \item A comparative analysis of Mamba, self-attention, LSTM, and Mamba-S4 for audio denoising.
    \item An efficient, structured pruning strategy using periodic calibration of GroupTaylor importance~\cite{molchanov2019importance}.
\end{enumerate}

\begin{figure}[t!]
  \centering
  \includegraphics[width=0.45\textwidth]{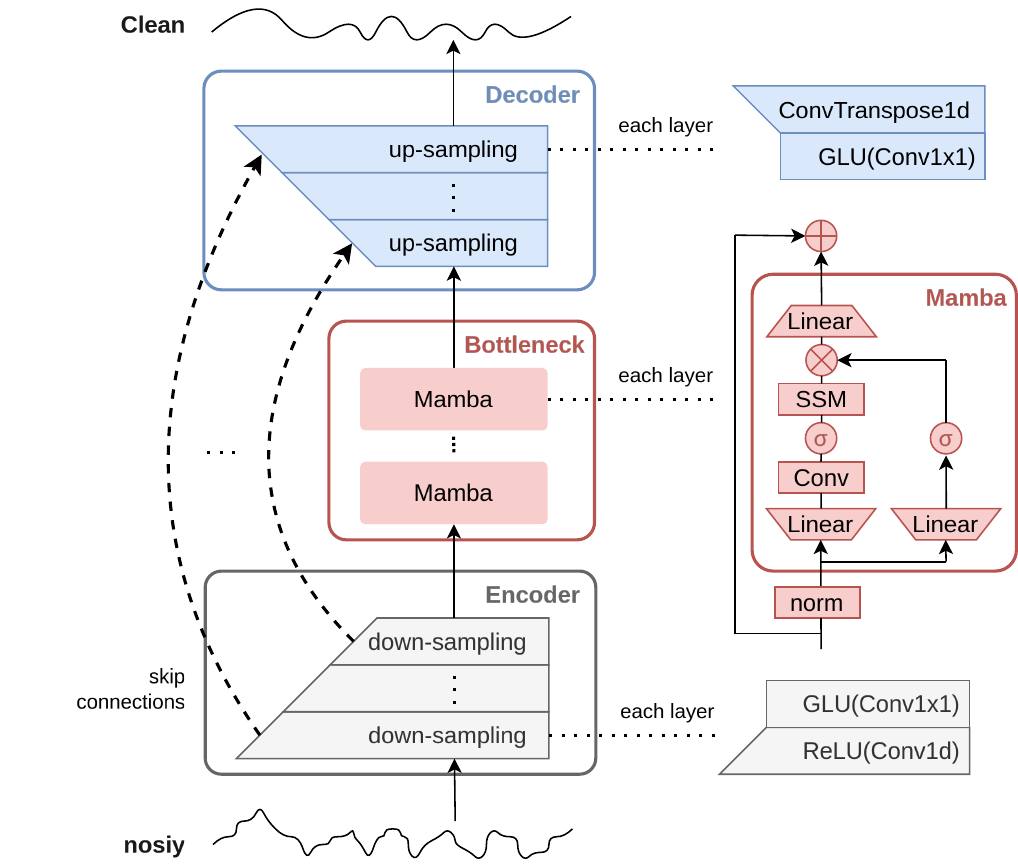}
  \caption{\texttt{CleanUMamba} architecture consisting of convolutional encoder-decoder with 3 sequential Mamba blocks in the bottleneck.}
  \label{fig:CleanUMamba}
\end{figure}

\section{Related Works}

Recent research has explored the application of Mamba to audio denoising, with several concurrent works emerging. Many of these studies focus on non-causal speech enhancement for offline settings, where the entire audio track is available for processing. Notable works include \texttt{SPMamba}~\cite{li2024spmamba}, \texttt{DPMamba}~\cite{jiang2024dual}, \texttt{Mamba-TasNet}~\cite{jiang2024speech}, and \texttt{SEMamba} \cite{chao2024investigation}, which build upon \texttt{TF-Gridnet}~\cite{wang2023tf}, \texttt{SlowFast}~\cite{cheng2025modulatingstatespacemodel}, \texttt{Dual-path RNN} \cite{luo2020dual}, \texttt{Conv-TasNet}~\cite{luo2019conv}, and \texttt{MP-SENet}~\cite{lu2023mp} respectively. These approaches incorporate bidirectional Mamba implementations for efficient global audio processing. \texttt{TRAMBA}~\cite{sui2024tramba} enhances a time domain U-Net Temporal \texttt{FiLM}~\cite{birnbaum2019temporal} by integrating Mamba in the bottleneck and using self-attention for feature-wise linear modulation. This network is applied to audio super-resolution and reconstruction from bone-conducting microphone and accelerometer data. For long-term streaming applications, \texttt{oSpatialNet-Mamba}~\cite{quan2024multichannel} extends \texttt{SpatialNet}~\cite{quan2024spatialnet} by replacing self-attention with masked self-attention, Retention, or Mamba. Operating in the time-frequency domain with 2-4 second window sizes, Mamba outperforms other variations. Zhang et al.~\cite{zhang2024mamba} conduct an ablation study comparing different backbones, including Mamba and two bidirectional Mamba implementations, in the time-frequency domain.

In contrast to these works, our research focuses on real-time causal speech enhancement in the time domain. By applying Mamba to the U-Net architecture, we reduce computational load compared to DPMamba and Mamba-Tasnet through processing in a lower-resolution latent space.


\section{Proposed Method}
\subsection{Problem Definition}
\begin{figure}[t!]
  \centering
  \includegraphics[width=0.45\textwidth]{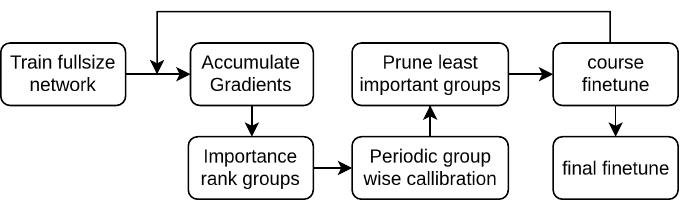}
  \caption{Pruning Pipeline}
  \label{fig:pruning_pipeline}
\end{figure}

Audio denoising aims to recover the clean speech signal \(x \in R^T\) from the noisy signal \(y = x + v\), where \(v\) represents zero-mean noise uncorrelated with \(x\). A causal model for denoising reconstructs the clean speech \(\hat{x}_t = f(y_{1:t}) \approx x_t\) using noisy samples up to time \(t\). In real-world applications, a slight look-ahead, such as a delay of 5-6 ms for hearing aids \cite{stone2008tolerable, roth2024effect}, or up to 200 ms for video calls \cite{ITU-T_G.114}, is acceptable. This work uses 16kHz audio with \texttt{CleanUMamba}, achieving an algorithmic delay of 48 ms and 12 ms for 8 and 6 encoder layers, respectively.

\subsection{Network Architecture}
\begin{table}[t!]
 \centering
 \caption{Multi-Head Attention (MHA), LSTM, Mamba, and Mamba S4 compared in the same size unit model and training conditions on the DNS no-reverb test set.}
 \label{tab:ablation_results}
 \begin{tabular}{l@{\hspace{0.5\tabcolsep}}|@{\hspace{0.5\tabcolsep}}r@{\hspace{0.5\tabcolsep}}r|l@{\hspace{0.5\tabcolsep}}l@{\hspace{0.5\tabcolsep}}l@{\hspace{0.5\tabcolsep}}l@{\hspace{0.5\tabcolsep}}l@{\hspace{0.5\tabcolsep}}l@{\hspace{0.5\tabcolsep}}}
 
 \toprule
 Model & Params & MACs & \begin{tabular}[c]{@{}l@{}}PESQ\\ (WB)\end{tabular} & \begin{tabular}[c]{@{}l@{}}PESQ\\ (NB)\end{tabular} & \begin{tabular}[c]{@{}l@{}}STOI\\ (\%)\end{tabular} & \begin{tabular}[c]{@{}l@{}}pred.\\ CSIG\end{tabular} & \begin{tabular}[c]{@{}l@{}}pred.\\ CBAK\end{tabular} & \begin{tabular}[c]{@{}l@{}}pred.\\ COVRL\end{tabular} \\ \hline

 Mamba & \underline{442K} & 468M & \textbf{2.42} & \textbf{2.95} & \textbf{95.1} & \textbf{3.98} & \textbf{3.25} & \textbf{3.21} \\
 Mamba S4 & 451K & 468M & 2.36 & 2.90 & \underline{94.9} & 3.93 & \underline{3.20} & 3.15 \\
 MHA & \underline{443K} & 470M & \underline{2.37} & \underline{2.92} & \underline{94.9} & \underline{3.94} & \underline{3.20} & \underline{3.16} \\
 LSTM & \underline{443K} & \underline{463M} & 2.32 & 2.88 & 94.7 & 3.90 & 3.19 & 3.12 \\ 
 \bottomrule
 \end{tabular}
\end{table}

Figure~\ref{fig:CleanUMamba} shows the architecture of \texttt{CleanUMamba}, consisting of an encoder-decoder structure with \(E\) layers. Each encoder layer employs a 1D convolution with a kernel size of 4 and stride of 2, followed by a \texttt{ReLU} activation and a 1x1 convolution with a \texttt{GLU} activation. These layers halve the temporal resolution and include bypass connections to their respective decoder layers. Decoder layers reverse this process using 1D transposed convolutions, doubling the temporal resolution. The channels per layer start at \(H=64\) and increase up to a maximum of \(768\).

In the bottleneck, three Mamba blocks perform sequence modeling on the latent representation, which has a model dimension of \(D=512\). Each Mamba block is preceded by layer normalization and a residual connection, while the inner dimension is set to \(I=2048\), corresponding to CleanUNet's fully connected layer output \cite{Kong_2022_CleanUNet}. The state-space model uses \(S=64\) channels.

\subsection{Pruning Pipeline}
Pruning aims to minimize the loss increase per pruned parameter. Let \( \mathcal{L}(\theta) \) represent the loss function of the network with parameters \(\theta\), and \(\Delta \mathcal{L}\) denote the change in loss due to pruning. The objective is:

\begin{equation}
\min_{\mathcal{S}} \frac{\Delta \mathcal{L}(\mathcal{S})}{|\mathcal{S}|}
\end{equation}

where \( \mathcal{S} \) is the set of parameters to prune, and \( |\mathcal{S}| \) is its size. Since exact loss sampling for every group is computationally infeasible, the Group Taylor importance metric \cite{molchanov2019importance} is used. It estimates the importance of parameter sets with both absolute and squared gradients:

\begin{equation}
    I_S = \sum_{s\in S} |g_s w_s|
    \label{eq:group_taylor_importance}
\end{equation}
\begin{equation}
    I_S = \sum_{s\in S} (g_s w_s)^2
    \label{eq:group_taylor_squared_importance}
\end{equation}
\begin{equation}
    I_S = \sum_{s\in S} |w_s|
    \label{eq:group_weight_importance}
\end{equation}

Here, \(g_s\) and \(w_s\) represent the gradient and weight for group \(s\). Gradients are accumulated via backpropagation. The pruning pipeline (Fig.~\ref{fig:pruning_pipeline}) follows a standard train-prune-finetune approach \cite{han2015learning}, accumulating micro-batch gradients \cite{piao2023enabling}. Pruning targets a percentage of groups, selecting those with the lowest Group Taylor importance, while ensuring compatibility with Mamba's causal convolutions by pruning in multiples of 8 channels.

Though effective, the Taylor importance metric may over- or under-penalize groups of varying sizes or depths. To address this, periodic calibration is applied by pruning 20\% of groups in a layer, measuring the loss increase, and adjusting the global metric accordingly. An exponential moving average filter smooths out any noise, preventing a single outlier from disproportionately affecting the pruning process.

\section{Experimental Setup}
\subsection{Evaluation Metrics}
\begin{figure}[ht!]
 \centering
 \begin{subfigure}[b]{0.4\textwidth}
     \centering
     \includegraphics[width=\textwidth]{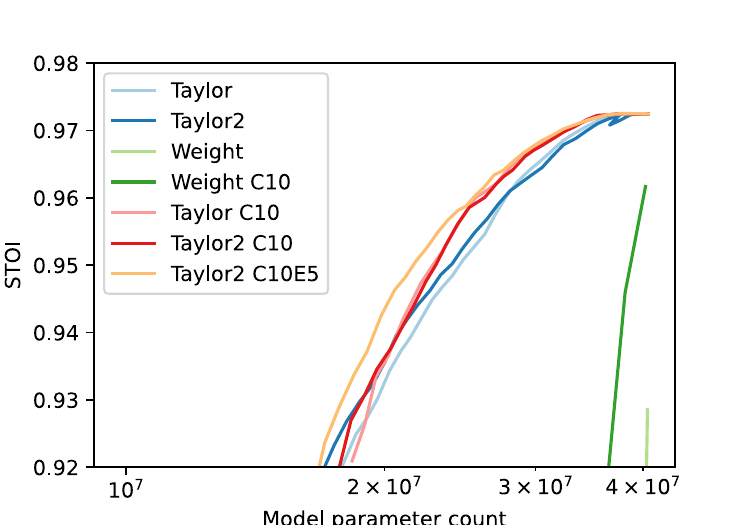}
     \caption{Model parameter count vs STOI}
 \end{subfigure}
 \begin{subfigure}[b]{0.4\textwidth}
     \centering
     \includegraphics[width=\textwidth]{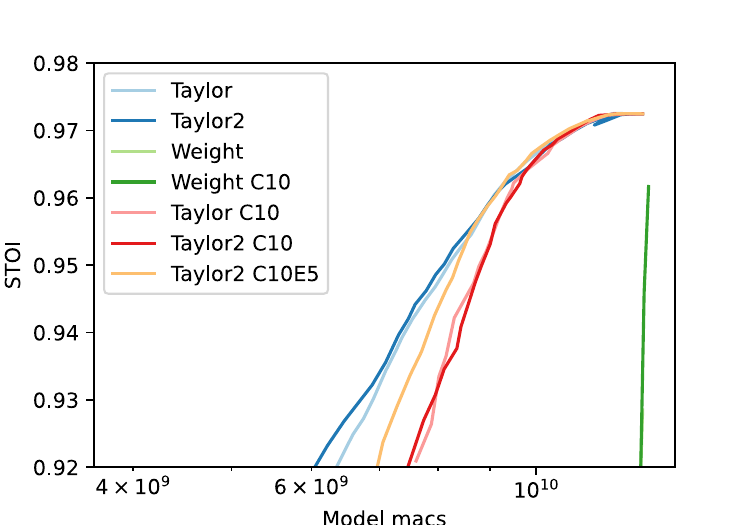}
     \caption{Model MACs vs STOI}
 \end{subfigure}
 \caption{Comparison of importance metrics for pruning \texttt{CleanUMamba} without fine-tuning.}
 \label{fig:no_finetune_experiments}
\end{figure}

We evaluated the model using several objective metrics: Perceptual Evaluation of Speech Quality (PESQ) \cite{rix2001perceptual}, Short-Time Objective Intelligibility (STOI) \cite{taal2011algorithm}, and the Mean Option Score (MOS) for signal distortion (CSIG), background noise intrusiveness (CBAK), and overall quality (COVL) \cite{hu2007evaluation}.

\subsection{Dataset}
The experiments used the Interspeech 2020 Deep Noise Suppression (DNS) dataset \cite{Dubey_2022_DNSC_Dataset}. This dataset includes speech from 2150 speakers and 65,000 noise clips, yielding 500 hours of training data, with signal-to-noise ratios (SNR) between -5 and 25 dB across 31 levels. During training, random 10-second crops were selected.

\subsection{Training Setup}
\label{section:LossFunction}
We employed a loss function combining L1 loss and multi-resolution short-time Fourier transform (STFT) loss \cite{yamamoto2020parallel}:

\begin{equation}
L(x, \hat{x}) = \|x-\hat{x}\| + STFT(x, \hat{x})
\end{equation}
{\small
\begin{equation}
\begin{split}
    STFT(x, \hat{x}) = \sum_{i=1}^{m} \left( \frac{\|s(x;\theta_i)-s(\hat{x};\theta_i)\|_F}{\|s(x;\theta_i)\|_F} + \frac{1}{T} \|\log{\frac{s(x;\theta_i)}{s(\hat{x};\theta_i)}}\right)
\end{split}
\end{equation}}

STFT loss was calculated for FFT bins \(\{512, 1024, 2048\}\), hop sizes \(\{50, 120, 240\}\), and window lengths \(\{240, 600, 1200\}\). Both full-band and high-frequency (4kHz-8kHz) STFT losses were tested \cite{Kong_2022_CleanUNet}.

The network was trained with the ADAM optimizer (learning rate 0.0002, \(\beta_1=0.9\), \(\beta_2=0.999\)) using a linear warm-up for the first 5\% of training followed by cosine decay. PyTorch AMP was applied for mixed-precision training. 

A Wiener filter baseline \cite{lim1978all}, implemented with the Pyroomacoustics library \cite{scheibler2018pyroomacoustics}, used an LPC order of 10, window size of 256, 2 iterations, smoothing alpha of 0.5, and threshold set for 16\% noise classification. The output was mixed with 50\% unfiltered audio to optimize quality.

\section{Experimental Results}
\subsection{Ablation Study of Mamba}

\begin{figure}[t!]
    \centering
    \includegraphics[width=0.4\textwidth]{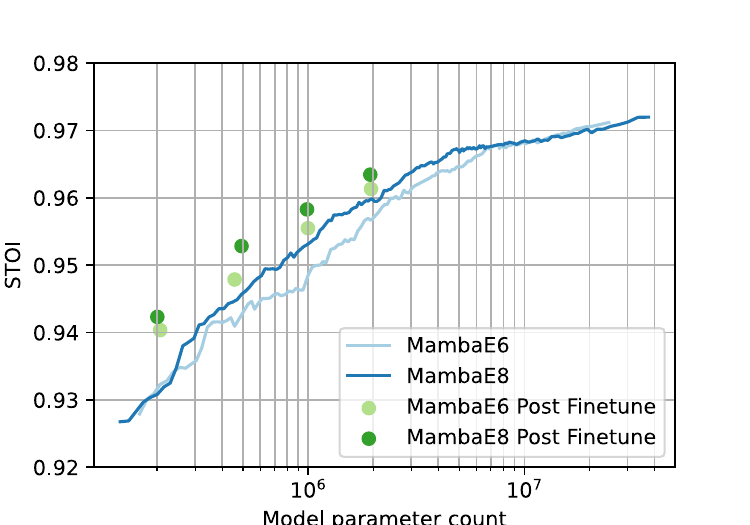}
    \caption{Pruning \texttt{CleanUMamba} with 6 and 8 encoder layers with fine-tuning: Model parameter count vs STOI}
    \label{fig:finetune_experiments}
\end{figure}
\begin{table*}[ht!]
    \centering
    \caption{Evaluation results for denoising on the DNS no-reverb testset.}
    \label{tab:CleanUMambaResultsDNS}
\begin{tabular}{l|rrr|llllll}
    \toprule
    Model & Params & MACs &  
    \begin{tabular}[c]{@{}l@{}}look\\ahead\end{tabular} & 
    \begin{tabular}[c]{@{}l@{}}PESQ\\ (WB)\end{tabular} & 
    \begin{tabular}[c]{@{}l@{}}PESQ\\ (NB)\end{tabular} & 
    \begin{tabular}[c]{@{}l@{}}STOI\\ (\%)\end{tabular} & 
    \begin{tabular}[c]{@{}l@{}}pred.\\ CSIG\end{tabular} & 
    \begin{tabular}[c]{@{}l@{}}pred.\\ CBAK\end{tabular} & 
    \begin{tabular}[c]{@{}l@{}}pred.\\ COVRL\end{tabular} \\ \hline

    raw noisy waveform & - & - & - & 1.585 & 2.164 & 91.6 & 3.091 & 2.539 & 2.304 \\
    wiener filter & - & - & 32ms & 1.645 & 2.237 & 91.6 & 3.099 & 2.558 & 2.336 \\
    DEMUCS \cite{Défossez_2020_demucs} & 33.53M & - & 40ms & 2.659 & 3.229 & 96.6 & 4.145 & 3.627 & 3.419 \\
    FullSubNet \cite{hao2021fullsubnet} & 5.6   M & - & 32ms & 2.777 & 3.305 & 96.1 & - & - & - \\

    CleanUNet N5 full \cite{Kong_2022_CleanUNet} & 46.07M & 13.66G & 48ms & 3.146 & 3.551 & 97.7 & 4.619 & 3.892 & 3.932 \\
    CleanUNet N5 high \cite{Kong_2022_CleanUNet} & 46.07M & 13.66G & 48ms & 3.011 & 3.460 & 97.3 & 4.513 & 3.812 & 3.800 \\

    CleanUNet N3 full \cite{Kong_2022_CleanUNet} & 39.77M & 13.18G & 48ms & 3.128 & 3.539 & 97.6 & - & - & - \\ 
    
    CleanUNet N3 high \cite{Kong_2022_CleanUNet} & 39.77M & 13.18G & 48ms & 3.006 & 3.453 & 97.3 & - & - & - \\

    \hline \hline
    CleanUMamba E8 full (ours) & 41.37M & 13.38G & 48ms & 3.067 & 3.507 & 97.4 & 4.502 & 3.870 & 3.829 \\
    CleanUMamba E8 high (ours)& 41.37M & 13.38G & 48ms & 3.017 & 3.471 & 97.2 & 4.456 & 3.840 & 3.775 \\ \hline
    \multirow{7}{*}{Pruned CleanUMamba E8 high (ours)}  & 14.90M & 6.08G & \multirow{7}{*}{48ms} & 2.910 & 3.397 & 97.0 & 4.393 & 3.742 & 3.682 \\
                                            & 6.00M & 3.87G &  & 2.888 & 3.359 & 96.9 & 4.380 & 3.711 & 3.661 \\
                                            & 3.22M & 2.09G &  & 2.746 & 3.253 & 96.4 & 4.272 & 3.599 & 3.530 \\
                                            & 1.94M & 1.97G  &  & 2.707 & 3.222 & 96.3 & 4.244 & 3.570 & 3.495 \\
                                            & 0.99M & 1.29G  &  & 2.558 & 3.102 & 95.8 & 4.125 & 3.466 & 3.356 \\
                                            & 492K & 807M  &  & 2.426 & 2.980 & 95.3 & 3.996 & 3.351 & 3.219 \\
                                            & 201K & 403M  &  & 2.189 & 2.745 & 94.2 & 3.784 & 3.163 & 2.986 \\
                                            \hline
    
    CleanUMamba E6 high (ours)& 27.21M & 13.48G & 12ms & 2.935 & 3.400 & 97.1 & 4.415 & 3.785 & 3.710 \\ \hline

\multirow{6}{*}{Pruned CleanUMamba E6 high (ours)}  
     & 13.50M & 7.49G  & \multirow{6}{*}{12ms}  & 2.855 & 3.346 & 96.9 & 4.366 & 3.710 & 3.637 \\
     & 7.31M & 4.97G  &  & 2.799 & 3.291 & 96.8 & 4.331 & 3.659 & 3.590 \\
     & 1.95M & 2.14G  &  & 2.602 & 3.128 & 96.1 & 4.171 & 3.499 & 3.402 \\
     & 1.00M & 1.36G  &  & 2.431 & 2.967 & 95.5 & 4.033 & 3.365 & 3.241 \\
     & 457K & 858M  &  & 2.237 & 2.796 & 94.8 & 3.855 & 3.218 & 3.048 \\
     & 207K & 483M  &  & 2.096 & 2.660 & 94.0 & 3.715 & 3.104 & 2.902 \\
    \bottomrule
\end{tabular}
\end{table*}
In Table~\ref{tab:ablation_results}, we compare the performance of Mamba in the bottleneck layer against Multi-head Attention, Mamba S4, and LSTM. All models use the same encoder-decoder architecture with \(E=8\) layers, starting with \(H=32\) channels in the first layer and \(H=64\) channels in subsequent layers. Each model employs a bottleneck with a model dimension of \(D=64\).

For the multi-head attention variant, 4 heads were used, each with a dimension of 16, followed by an \texttt{MLP} expanding to \(d\_inner=128\), in line with the architecture in \cite{Kong_2022_CleanUNet}, but with fewer parameters. The Mamba model uses an inner dimension of \(d\_inner=128\) and a state size \(S=16\), while Mamba S4 uses the same configuration but replaces the selective state space with the S4 state space \cite{Gu_2021_S4}. The LSTM model consists of 3 layers with a hidden size of \(64\).

All models were trained on the DNS dataset for 1M iterations with a batch size of 16. The \texttt{ADAMW} optimizer was employed, with a weight decay of 0.1, using the full STFT loss. Autocast was disabled for Mamba S4 due to training errors.

\subsection{CleanUMamba}
In Table \ref{tab:CleanUMambaResultsDNS}, we present the results of training \texttt{CleanUMamba} with the same procedure as \texttt{CleanUNet}, using \(N=3\) bottleneck layers and encoder depths of \(E=6\) and \(E=8\). When trained with high-band loss, \texttt{CleanUMamba} outperforms both the same-size and larger \texttt{CleanUNet} models. However, under full-band loss, \texttt{CleanUMamba} slightly underperforms compared to \texttt{CleanUNet}.

The model with \(E=6\) performs marginally worse than the \(E=8\) version but has fewer than 75\% of the parameters. Despite this, the number of multiply accumulates per second (MAC/s) remains comparable due to the longer input to the bottleneck layer.

With \(E=8\), most of the computations occur in the encoder and decoder, leading to a doubling of MAC/s after 18 minutes, making Mamba's fixed state size less significant. Reducing the encoder depth to \(E=6\) quadruples the bottleneck sequence length and reduces MAC/s in the encoder-decoder. As a result, after just 80 seconds, MAC/s with multi-head attention doubles compared to Mamba.

\subsection{Different Importance Metrics without Fine-Tuning}
To assess the effectiveness of different importance metrics, the \texttt{CleanUMamba} network trained with high loss was pruned without fine-tuning, as shown in Figure \ref{fig:no_finetune_experiments}. At each pruning step, 24 of the least important groups were removed, with 128 audio samples used for gradient accumulation in the Taylor importance metric. The calibrated runs, denoted \(C10\), recalibrate the importance metric every 10 pruning steps, and the exponential moving average, \(E5\), applies a smoothing factor of 0.5.

Results indicate that the Group Taylor importance metric effectively reflects global importance. The squared Taylor loss slightly outperforms the absolute Taylor loss at higher pruning levels. In contrast, weight magnitude performs poorly as a global importance metric, with significant performance drops after pruning the first few layers.

With calibration, Taylor loss initially shows better results, but stability decreases over time. Filtering the calibration improves quality per parameter, while uncalibrated Taylor loss consistently delivers superior quality per MAC/s.

\subsection{Pruning with Fine-Tuning}
For pruning with fine-tuning, \texttt{CleanUMamba} models with \(E=6\) and \(E=8\) encoder layers pre-trained using high loss were pruned using squared Taylor importance. The metric was recalibrated every 20 steps with a smoothing factor of 0.5. At each step, gradients were accumulated over 128 samples, and the model was fine-tuned on 40,960 training samples every 5 steps. The importance pruned was limited to 3e-13, with 4 channels pruned per step once this limit was reached.

However, both E6 and E8 models became under-trained below 6M and 4M parameters, respectively. To address this, models at 200K, 500K, 1M, and 2M parameters were fine-tuned for an additional 100K iterations with a batch size of 16.

Figure \ref{fig:finetune_experiments} shows the STOI scores during pruning, and Table \ref{tab:CleanUMambaResultsDNS} presents a subset of the full evaluation. At 3.22M parameters, \texttt{CleanUMamba} matches the performance of DEMUCS with 8$\times$ fewer parameters and nearly matches FullSubNet with 2$\times$ fewer parameters.

\section{Conclusion}
In this study, we introduce \texttt{CleanUMamba}, a real-time speech denoising model that operates in the waveform domain and investigate its size reduction through pruning. We evaluated the model on the Interspeech 2020 Deep Noise Suppression challenge and compared it to the self-attention-based \texttt{CleanUNet}. While both models perform similarly with an encoder depth of 8, Mamba's linear time complexity becomes more advantageous with a reduced depth of 6, achieving a 12\,ms latency. Our pruning pipeline further reduces model size, achieving performance comparable to LSTM-based DEMUCs with 8$\times$ fewer parameters.

\bibliographystyle{IEEEtran}
\bibliography{refs.bib}

\begin{thebibliography}{10}
\providecommand{\url}[1]{#1}
\csname url@samestyle\endcsname
\providecommand{\newblock}{\relax}
\providecommand{\bibinfo}[2]{#2}
\providecommand{\BIBentrySTDinterwordspacing}{\spaceskip=0pt\relax}
\providecommand{\BIBentryALTinterwordstretchfactor}{4}
\providecommand{\BIBentryALTinterwordspacing}{\spaceskip=\fontdimen2\font plus
\BIBentryALTinterwordstretchfactor\fontdimen3\font minus \fontdimen4\font\relax}
\providecommand{\BIBforeignlanguage}[2]{{%
\expandafter\ifx\csname l@#1\endcsname\relax
\typeout{** WARNING: IEEEtran.bst: No hyphenation pattern has been}%
\typeout{** loaded for the language `#1'. Using the pattern for}%
\typeout{** the default language instead.}%
\else
\language=\csname l@#1\endcsname
\fi
#2}}
\providecommand{\BIBdecl}{\relax}
\BIBdecl

\bibitem{boll1979suppression}
S.~Boll, ``Suppression of acoustic noise in speech using spectral subtraction,'' \emph{IEEE Transactions on acoustics, speech, and signal processing}, vol.~27, no.~2, pp. 113--120, 1979.

\bibitem{lim1979enhancement}
J.~S. Lim and A.~V. Oppenheim, ``Enhancement and bandwidth compression of noisy speech,'' \emph{Proceedings of the IEEE}, vol.~67, no.~12, pp. 1586--1604, 1979.

\bibitem{pascual2017segan}
S.~Pascual, A.~Bonafonte, and J.~Serra, ``Segan: Speech enhancement generative adversarial network,'' \emph{arXiv preprint arXiv:1703.09452}, 2017.

\bibitem{luo2019conv}
Y.~Luo and N.~Mesgarani, ``Conv-tasnet: Surpassing ideal time--frequency magnitude masking for speech separation,'' \emph{IEEE/ACM transactions on audio, speech, and language processing}, vol.~27, no.~8, pp. 1256--1266, 2019.

\bibitem{stoller2018wave}
D.~Stoller, S.~Ewert, and S.~Dixon, ``Wave-u-net: A multi-scale neural network for end-to-end audio source separation,'' \emph{arXiv preprint arXiv:1806.03185}, 2018.

\bibitem{pandey2019tcnn}
A.~Pandey and D.~Wang, ``Tcnn: Temporal convolutional neural network for real-time speech enhancement in the time domain,'' in \emph{ICASSP 2019-2019 IEEE International Conference on Acoustics, Speech and Signal Processing (ICASSP)}.\hskip 1em plus 0.5em minus 0.4em\relax IEEE, 2019, pp. 6875--6879.

\bibitem{Défossez_2020_demucs}
A.~Défossez, A.~Défossez, G.~Synnaeve, G.~Synnaeve, Y.~Adi, and Y.~Adi, ``Real time speech enhancement in the waveform domain.'' \emph{Interspeech}, 2020.

\bibitem{braun2021towards}
S.~Braun, H.~Gamper, C.~K. Reddy, and I.~Tashev, ``Towards efficient models for real-time deep noise suppression,'' in \emph{ICASSP 2021-2021 IEEE International Conference on Acoustics, Speech and Signal Processing (ICASSP)}.\hskip 1em plus 0.5em minus 0.4em\relax IEEE, 2021, pp. 656--660.

\bibitem{cheng24_interspeech}
L.~Cheng, A.~Pandey, B.~Xu, T.~Delbruck, and S.-C. Liu, ``Dynamic gated recurrent neural network for compute-efficient speech enhancement,'' in \emph{Interspeech 2024}, 2024, pp. 677--681.

\bibitem{Kong_2022_CleanUNet}
Z.~Kong, W.~Ping, A.~Dantrey, and B.~Catanzaro, ``Speech denoising in the waveform domain with self-attention,'' \emph{IEEE International Conference on Acoustics, Speech, and Signal Processing}, 2022.

\bibitem{wang2023tf}
Z.-Q. Wang, S.~Cornell, S.~Choi, Y.~Lee, B.-Y. Kim, and S.~Watanabe, ``Tf-gridnet: Making time-frequency domain models great again for monaural speaker separation,'' in \emph{ICASSP 2023-2023 IEEE International Conference on Acoustics, Speech and Signal Processing (ICASSP)}.\hskip 1em plus 0.5em minus 0.4em\relax IEEE, 2023, pp. 1--5.

\bibitem{Gu_2023_Mamba}
A.~Gu and T.~Dao, ``Mamba: Linear-time sequence modeling with selective state spaces,'' \emph{arXiv.org}, 2023.

\bibitem{molchanov2019importance}
P.~Molchanov, A.~Mallya, S.~Tyree, I.~Frosio, and J.~Kautz, ``Importance estimation for neural network pruning,'' in \emph{Proceedings of the IEEE/CVF conference on computer vision and pattern recognition}, 2019, pp. 11\,264--11\,272.

\bibitem{li2024spmamba}
K.~Li and G.~Chen, ``Spmamba: State-space model is all you need in speech separation,'' \emph{arXiv preprint arXiv:2404.02063}, 2024.

\bibitem{jiang2024dual}
X.~Jiang, C.~Han, and N.~Mesgarani, ``Dual-path mamba: Short and long-term bidirectional selective structured state space models for speech separation,'' \emph{arXiv preprint arXiv:2403.18257}, 2024.

\bibitem{jiang2024speech}
X.~Jiang, Y.~A. Li, A.~N. Florea, C.~Han, and N.~Mesgarani, ``Speech slytherin: Examining the performance and efficiency of mamba for speech separation, recognition, and synthesis,'' \emph{arXiv preprint arXiv:2407.09732}, 2024.

\bibitem{chao2024investigation}
R.~Chao, W.-H. Cheng, M.~La~Quatra, S.~M. Siniscalchi, C.-H.~H. Yang, S.-W. Fu, and Y.~Tsao, ``An investigation of incorporating mamba for speech enhancement,'' \emph{arXiv preprint arXiv:2405.06573}, 2024.

\bibitem{cheng2025modulatingstatespacemodel}
\BIBentryALTinterwordspacing
L.~Cheng, A.~Pandey, B.~Xu, T.~Delbruck, V.~K. Ithapu, and S.-C. Liu, ``Modulating state space model with slowfast framework for compute-efficient ultra low-latency speech enhancement,'' 2025. [Online]. Available: \url{https://arxiv.org/abs/2411.02019}
\BIBentrySTDinterwordspacing

\bibitem{luo2020dual}
Y.~Luo, Z.~Chen, and T.~Yoshioka, ``Dual-path rnn: efficient long sequence modeling for time-domain single-channel speech separation,'' in \emph{ICASSP 2020-2020 IEEE International Conference on Acoustics, Speech and Signal Processing (ICASSP)}.\hskip 1em plus 0.5em minus 0.4em\relax IEEE, 2020, pp. 46--50.

\bibitem{lu2023mp}
Y.-X. Lu, Y.~Ai, and Z.-H. Ling, ``Mp-senet: A speech enhancement model with parallel denoising of magnitude and phase spectra,'' \emph{arXiv preprint arXiv:2305.13686}, 2023.

\bibitem{sui2024tramba}
Y.~Sui, M.~Zhao, J.~Xia, X.~Jiang, and S.~Xia, ``Tramba: A hybrid transformer and mamba architecture for practical audio and bone conduction speech super resolution and enhancement on mobile and wearable platforms,'' \emph{arXiv preprint arXiv:2405.01242}, 2024.

\bibitem{birnbaum2019temporal}
S.~Birnbaum, V.~Kuleshov, Z.~Enam, P.~W.~W. Koh, and S.~Ermon, ``Temporal film: Capturing long-range sequence dependencies with feature-wise modulations.'' \emph{Advances in Neural Information Processing Systems}, vol.~32, 2019.

\bibitem{quan2024multichannel}
C.~Quan and X.~Li, ``Multichannel long-term streaming neural speech enhancement for static and moving speakers,'' \emph{IEEE Signal Processing Letters}, vol.~31, pp. 2295--2299, 2024.

\bibitem{quan2024spatialnet}
\BIBentryALTinterwordspacing
------, ``Spatialnet: Extensively learning spatial information for multichannel joint speech separation, denoising and dereverberation,'' \emph{IEEE/ACM Trans. Audio, Speech and Lang. Proc.}, vol.~32, p. 1310–1323, Feb. 2024. [Online]. Available: \url{https://doi-org.tudelft.idm.oclc.org/10.1109/TASLP.2024.3357036}
\BIBentrySTDinterwordspacing

\bibitem{zhang2024mamba}
X.~Zhang, Q.~Zhang, H.~Liu, T.~Xiao, X.~Qian, B.~Ahmed, E.~Ambikairajah, H.~Li, and J.~Epps, ``Mamba in speech: Towards an alternative to self-attention,'' \emph{arXiv preprint arXiv:2405.12609}, 2024.

\bibitem{stone2008tolerable}
M.~A. Stone, B.~C. Moore, K.~Meisenbacher, and R.~P. Derleth, ``Tolerable hearing aid delays. v. estimation of limits for open canal fittings,'' \emph{Ear and hearing}, vol.~29, no.~4, pp. 601--617, 2008.

\bibitem{roth2024effect}
S.~Roth, F.-U. M{\"u}ller, J.~Angermeier, W.~Hemmert, and S.~Zirn, ``Effect of a processing delay between direct and delayed sound in simulated open fit hearing aids on speech intelligibility in noise,'' \emph{Frontiers in Neuroscience}, vol.~17, p. 1257720, 2024.

\bibitem{ITU-T_G.114}
{International Telecommunication Union}, ``Itu-t recommendation g.114: One-way transmission time,'' Telecommunication Standardization Sector of ITU, Tech. Rep. G.114, 5 2003.

\bibitem{han2015learning}
S.~Han, J.~Pool, J.~Tran, and W.~Dally, ``Learning both weights and connections for efficient neural network,'' \emph{Advances in neural information processing systems}, vol.~28, 2015.

\bibitem{piao2023enabling}
X.~Piao, D.~Synn, J.~Park, and J.-K. Kim, ``Enabling large batch size training for dnn models beyond the memory limit while maintaining performance,'' \emph{IEEE Access}, 2023.

\bibitem{rix2001perceptual}
A.~W. Rix, J.~G. Beerends, M.~P. Hollier, and A.~P. Hekstra, ``Perceptual evaluation of speech quality (pesq)-a new method for speech quality assessment of telephone networks and codecs,'' in \emph{2001 IEEE international conference on acoustics, speech, and signal processing. Proceedings (Cat. No. 01CH37221)}, vol.~2.\hskip 1em plus 0.5em minus 0.4em\relax IEEE, 2001, pp. 749--752.

\bibitem{taal2011algorithm}
C.~H. Taal, R.~C. Hendriks, R.~Heusdens, and J.~Jensen, ``An algorithm for intelligibility prediction of time--frequency weighted noisy speech,'' \emph{IEEE Transactions on audio, speech, and language processing}, vol.~19, no.~7, pp. 2125--2136, 2011.

\bibitem{hu2007evaluation}
Y.~Hu and P.~C. Loizou, ``Evaluation of objective quality measures for speech enhancement,'' \emph{IEEE Transactions on audio, speech, and language processing}, vol.~16, no.~1, pp. 229--238, 2007.

\bibitem{Dubey_2022_DNSC_Dataset}
H.~Dubey, H.~Dubey, H.~Dubey, H.~Dubey, V.~Gopal, V.~Gopal, V.~Gopal, V.~Gopal, R.~Cutler, R.~Cutler, R.~Cutler, R.~Cutler, A.~Aazami, A.~Aazami, A.~Aazami, A.~Aazami, S.~Matusevych, S.~Matusevych, S.~Matusevych, S.~Matusevych, S.~Braun, S.~Braun, S.~Braun, S.~Braun, Şefik Emre~Eskimez, S.~E. Eskimez, S.~E. Eskimez, S.~E. Eskimez, M.~Thakker, M.~Thakker, M.~Thakker, M.~Thakker, T.~Yoshioka, T.~Yoshioka, T.~Yoshioka, T.~Yoshioka, H.~Gamper, H.~Gamper, H.~Gamper, H.~Gamper, R.~Aichner, R.~Aichner, R.~Aichner, and R.~Aichner, ``Icassp 2022 deep noise suppression challenge,'' \emph{IEEE International Conference on Acoustics, Speech, and Signal Processing}, 2022.

\bibitem{yamamoto2020parallel}
R.~Yamamoto, E.~Song, and J.-M. Kim, ``Parallel wavegan: A fast waveform generation model based on generative adversarial networks with multi-resolution spectrogram,'' in \emph{ICASSP 2020-2020 IEEE International Conference on Acoustics, Speech and Signal Processing (ICASSP)}.\hskip 1em plus 0.5em minus 0.4em\relax IEEE, 2020, pp. 6199--6203.

\bibitem{lim1978all}
J.~Lim and A.~Oppenheim, ``All-pole modeling of degraded speech,'' \emph{IEEE Transactions on Acoustics, Speech, and Signal Processing}, vol.~26, no.~3, pp. 197--210, 1978.

\bibitem{scheibler2018pyroomacoustics}
R.~Scheibler, E.~Bezzam, and I.~Dokmani{\'c}, ``Pyroomacoustics: A python package for audio room simulation and array processing algorithms,'' in \emph{2018 IEEE international conference on acoustics, speech and signal processing (ICASSP)}.\hskip 1em plus 0.5em minus 0.4em\relax IEEE, 2018, pp. 351--355.

\bibitem{hao2021fullsubnet}
X.~Hao, X.~Su, R.~Horaud, and X.~Li, ``Fullsubnet: A full-band and sub-band fusion model for real-time single-channel speech enhancement,'' in \emph{ICASSP 2021-2021 IEEE International Conference on Acoustics, Speech and Signal Processing (ICASSP)}.\hskip 1em plus 0.5em minus 0.4em\relax IEEE, 2021, pp. 6633--6637.

\bibitem{Gu_2021_S4}
A.~Gu, K.~Goel, and C.~R'e, ``Efficiently modeling long sequences with structured state spaces,'' \emph{International Conference on Learning Representations}, 2021.

\end{thebibliography}
\end{document}